\documentclass[aip,apl,preprint]{revtex4-2}

\usepackage{amsmath}
\usepackage{graphicx}
\usepackage{dcolumn}	
\usepackage{hyperref}
\usepackage[T1]{fontenc}
\usepackage{textcomp}
\usepackage{bm}

\begin{document}

\title{Electrical spin-wave spectroscopy in nanoscale waveguides with nonuniform magnetization}
\author{Giacomo Talmelli}
\affiliation{Imec, B-3001 Leuven, Belgium}
\affiliation{KU Leuven, Departement Materiaalkunde, 3001 Leuven, Belgium}
\author{Daniele Narducci}
\affiliation{Imec, B-3001 Leuven, Belgium}
\affiliation{KU Leuven, Departement Materiaalkunde, 3001 Leuven, Belgium}
\affiliation{Universit\`a degli Studi di Roma ``La Sapienza'', Dipartimento di Ingegneria dell'Informazione, Elettronica e delle Telecomunicazioni, 00184 Roma, Italy}
\author{Frederic Vanderveken}
\affiliation{Imec, B-3001 Leuven, Belgium}
\affiliation{KU Leuven, Departement Materiaalkunde, 3001 Leuven, Belgium}
\author{Marc Heyns}
\affiliation{Imec, B-3001 Leuven, Belgium}
\affiliation{KU Leuven, Departement Materiaalkunde, 3001 Leuven, Belgium}
\author{Fernanda Irrera}
\affiliation{Universit\`a degli Studi di Roma ``La Sapienza'', Dipartimento di Ingegneria dell'Informazione, Elettronica e delle Telecomunicazioni, 00184 Roma, Italy}
\author{Inge Asselberghs}
\author{Iuliana P. Radu}
\author{Christoph Adelmann}
\affiliation{Imec, B-3001 Leuven, Belgium}
\author{Florin Ciubotaru}
\email[Author to whom correspondence should be addressed. E-mail: ]{Florin.Ciubotaru@imec.be}
\affiliation{Imec, B-3001 Leuven, Belgium}

\begin{abstract}
Spin waves modes in magnetic waveguides with width down to 320 nm have been studied by electrical propagating spin-wave spectroscopy and micromagnetic simulations for both longitudinal and transverse magnetic bias fields. For longitudinal bias fields, a 1.3 GHz wide spin-wave band was observed in agreement with analytical dispersion relations for uniform magnetization. However, transverse bias field led to several distinct bands, corresponding to different quantized width modes, with both negative and positive slopes. Micromagnetic simulations showed that, in this geometry, the magnetization was nonuniform and tilted due to the strong shape anisotropy of the waveguides. Simulations of the quantized spin-wave modes in such nonuniformly magnetized waveguides resulted in spin wave dispersion relations in good agreement with the experiments.
\end{abstract}

\maketitle

Spin waves are collective excitations of the magnetization in ferromagnetic materials with typical frequencies in the GHz and wavelengths in the nm to \textmu{}m ranges. Due to their low intrinsic energies, they have recently received increasing interest for ultralow power spintronic  applications,\cite{Dieny_2020,Zografos_2015} in particular in Boolean logic circuits\cite{Khitun_2011, Chumak_2015, Radu_2015, Mahmoud_2020} as well as analog\cite{Csaba_2014} or neuromorphic computing\cite{Papp_2020} concepts. Recently, several of the proposed device concepts have been realized experimentally.\cite{Fischer_2017, Kanazawa_2017,Balynsky_2018,Wang_2020,Talmelli_2020}

To be competitive with conventional electronics, the logic element density in a spintronic circuit must be high and therefore spin-wave devices must be miniaturized to nanoscale dimensions. At the nanoscale, the properties of spin waves are strongly affected by quantization and pinning conditions at the device edges. Spin-wave devices are commonly operated under transverse applied bias fields. However, in this configuration, the magnitude of the shape anisotropy field in nanoscale magnetic devices (in particular in long waveguides) can become comparable to typical bias fields (on the order of hundreds of mT), which leads to a nonuniform magnetization orientation inside the structure. This has a strong impact on the relative excitation efficiency of different quantized spin-wave modes as well as on their propagation.\cite{Vanderveken_2020}
Confined spin waves in uniformly magnetized nanoscale structures have been studied by micromagnetic simulations and imaged by optical and x-ray-based techniques,\cite{Wang_2019,Heinz_2020,Trager_2020} revealing the formation of quantized width modes. However, all-electrical experiments\cite{Bailleul_2003, Ciubotaru_2016, Collet_2017, Bhaskar_2020} at the nanoscale are still lacking due to challenges to detect spin-wave signals in reduced magnetic volumes.

To date, little work has been dedicated to scaled waveguides with nonuniform magnetization. In this letter we report an all-electrical study of spin waves in magnetic waveguides with widths down to 320 nm. We show that the nonuniformity of the anisotropy field strongly affects spin wave transmission spectra for transverse magnetic bias fields whereas the effects are much less pronounced for longitudinal bias fields. The experimental results are in good agreement with micromagnetic simulations.

 \begin{figure}[t]
	\includegraphics[width=8.5cm]{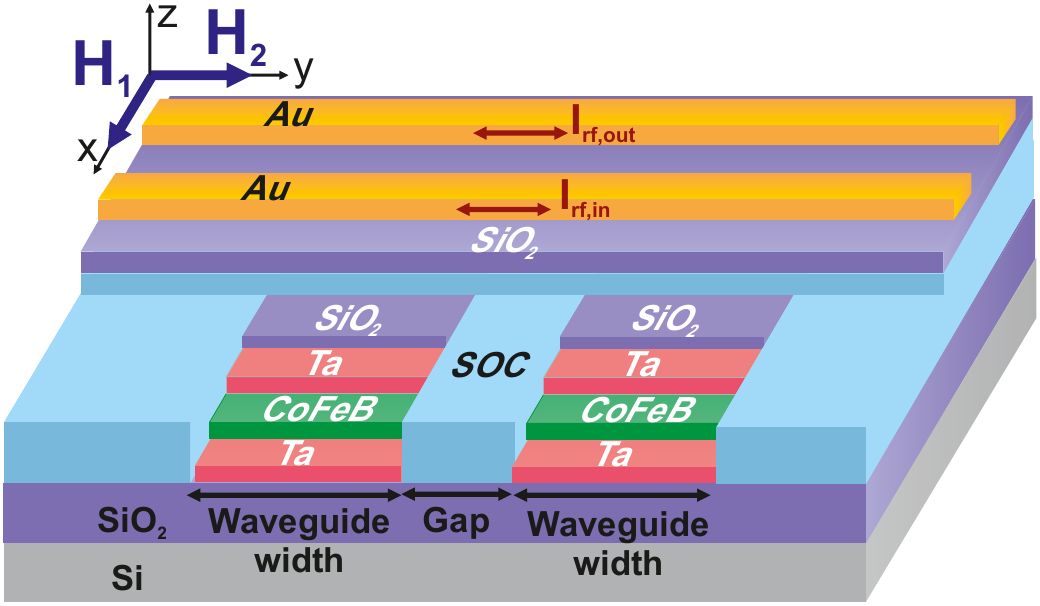}
	\caption{Schematic of a studied device including two parallel magnetic waveguides and two inductive antennas to generate and detect spin waves. }
	\label{Fig_Schematic} 
\end{figure}

The studied structures consisted of multiple parallel Ta/Co$_{40}$Fe$_{40}$B$_{20}$/Ta ferromagnetic waveguides of 50 \textmu{}m length. The number of waveguides was chosen so that the total magnetic width was about 5 \textmu{}m to ensure sufficient electrical read out signals.\cite{Bailleul_2003,Lassalle_2011, Collet_2017} Details of the sample fabrication can be found in Ref.~\onlinecite{Talmelli_2020}. The saturation magnetization of Co$_{40}$Fe$_{40}$B$_{20}$ was $M_{s}=1.36$ MA/m, determined by vibrating-sample magnetometry and ferromagnetic resonance (FMR). The FMR experiments further found a Gilbert damping parameter of $\alpha=4.3\times 10^{-3}$ and a Land\'e factor of $g = 2.07$.\cite{Talmelli_2020,Liu_2011} Spin waves were excited and detected by 500 nm wide Au inductive antennas, electrically connected to coplanar microwave waveguides. A schematic of the device structure is shown in Fig.~\ref{Fig_Schematic}.

\begin{figure}[t]
	\includegraphics[width=8.5cm]{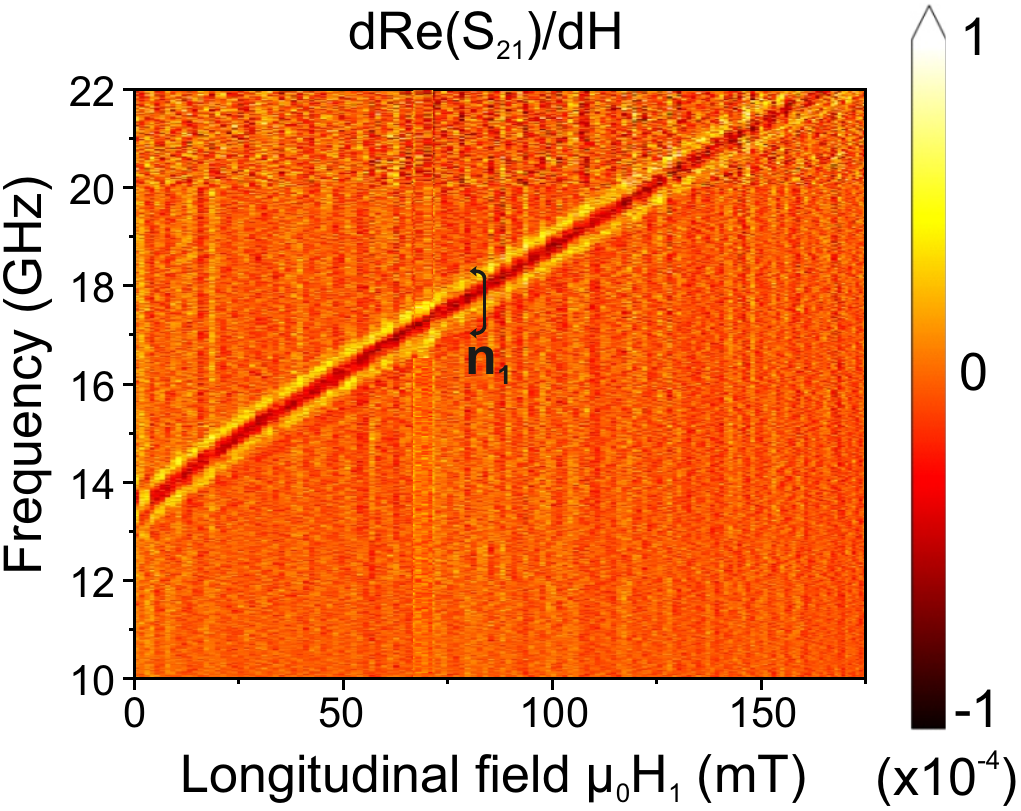}
	\caption{Bias-field derivative of the real part of the $S_{21}$ microwave transmission parameter corresponding to spin-wave propagation in 20 parallel 320 nm wide waveguides (separated by 180 nm gaps) for longitudinal bias field $H_{1}$. }
	\label{Fig_2DMaps} 
\end{figure}

Electrical measurements of the devices were performed using a Keysight E8363B network analyzer (output RF power 0.25 mW). Both the reflected ($S_{11}$) and transmitted ($S_{21}$) power was recorded \emph{vs.} magnetic bias field and frequency. Figure~\ref{Fig_2DMaps} shows the bias-field derivative of $S_{21}$ for a device including 20 waveguides (320 nm width, 180 nm gap) and longitudinal bias fields $H_1$ along the $x$-direction (\emph{cf.} Fig.~\ref{Fig_Schematic}). In this geometry, spin waves are of the backward volume type.\cite{Kalinikos_1986} The data indicate a spin-wave transmission band between 12.5 and 13.8~GHz close to zero bias field, which progressively shifts to higher frequencies with increasing $H_{1}$. The strong shape anisotropy leads to uniform magnetization in the waveguides even at very low magnetic fields, leading to the observation of the spin waves at magnetic fields close to zero. The spin-wave frequency bandwidth of about 1.3~GHz is in agreement with the spin-wave dispersion relations \cite{Kalinikos_1986} and is limited by the FMR frequency and the wavenumber spectrum determined by the antenna width.\cite{Ciubotaru_2016} Considering the bias field and the lateral confinement in the waveguide, analytical calculations and micromagnetic simulations (not shown) indicate that the measured signal corresponds to the propagation of the first-order quantized width mode $n_{1}$. The frequency band of the third-order width mode $n_{3}$ is expected about 7~GHz above the $n_{1}$ band and was not visible in the experiment. Note that the antenna can excite only odd width modes in this configuration.\cite{Demidov_2015}

\begin{figure}[t]
	\includegraphics[width=9.5 cm]{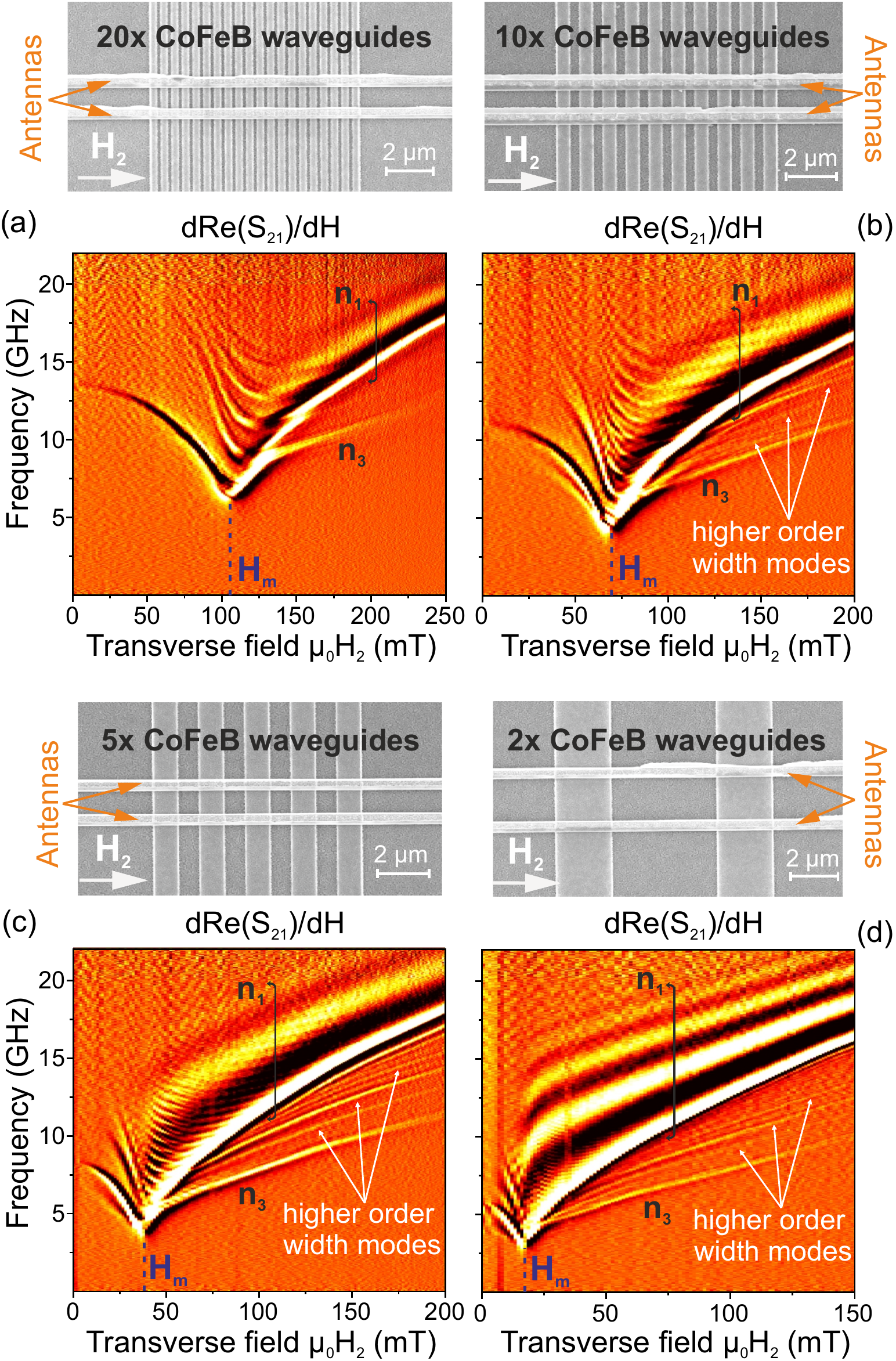}
	\caption{Scanning electron micrographs and bias-field derivatives of the real part of the transmitted power ($S_{21}$) for four different devices under transverse applied bias fields. The device parameters were: (a) 20 $\times$ 320 nm wide waveguides with 180 nm gaps, (b) 10 $\times$ 600 nm wide waveguides with 400 nm gaps, (c) 5 $\times$ 1.1 $\mu m$ wide waveguides with 0.9 $\mu m$ gaps, and (d) 2 $\times$ 2.5 $\mu m$ wide waveguides with 5 $\mu m$ gaps. The color scale is the same as in Fig.~\ref{Fig_2DMaps}}
	\label{DE_Exp}
\end{figure}

The behavior was much more complex for transverse applied bias fields $H_{2}$ in the Damon-Eshbach geometry. Experimental field--frequency maps of the bias-field derivative of $S_{21}$ are shown in Fig.~\ref{DE_Exp} for sets of waveguides with different widths between 320 nm and 2.5 \textmu{}m. The data reveal several spin-wave branches with positive and negative slopes depending on the bias field magnitude. A feature of all maps is a frequency minimum at a characteristic bias field $H_m$ that increases with decreasing waveguide width. The projection of the signal to $H_{2}=0$ leads to a frequency of $\sim13.8$~GHz for the narrowest waveguides (320 nm width, Fig.~\ref{DE_Exp}(a)), in agreement with the FMR frequency found in the backward volume configuration at zero bias field (\emph{cf.} Fig.~\ref{Fig_2DMaps}). This suggests that the strong shape anisotropy of the waveguides forces the magnetization along the waveguides at small transverse $H_{2}$. For increasing $H_{2}$, the magnetization progressively rotates towards the $y$-direction, leading to a positive slope of the spin-wave dispersion (positive group velocity) after a critical field $H_m$.

\begin{figure}[t]
\includegraphics[width=8cm]{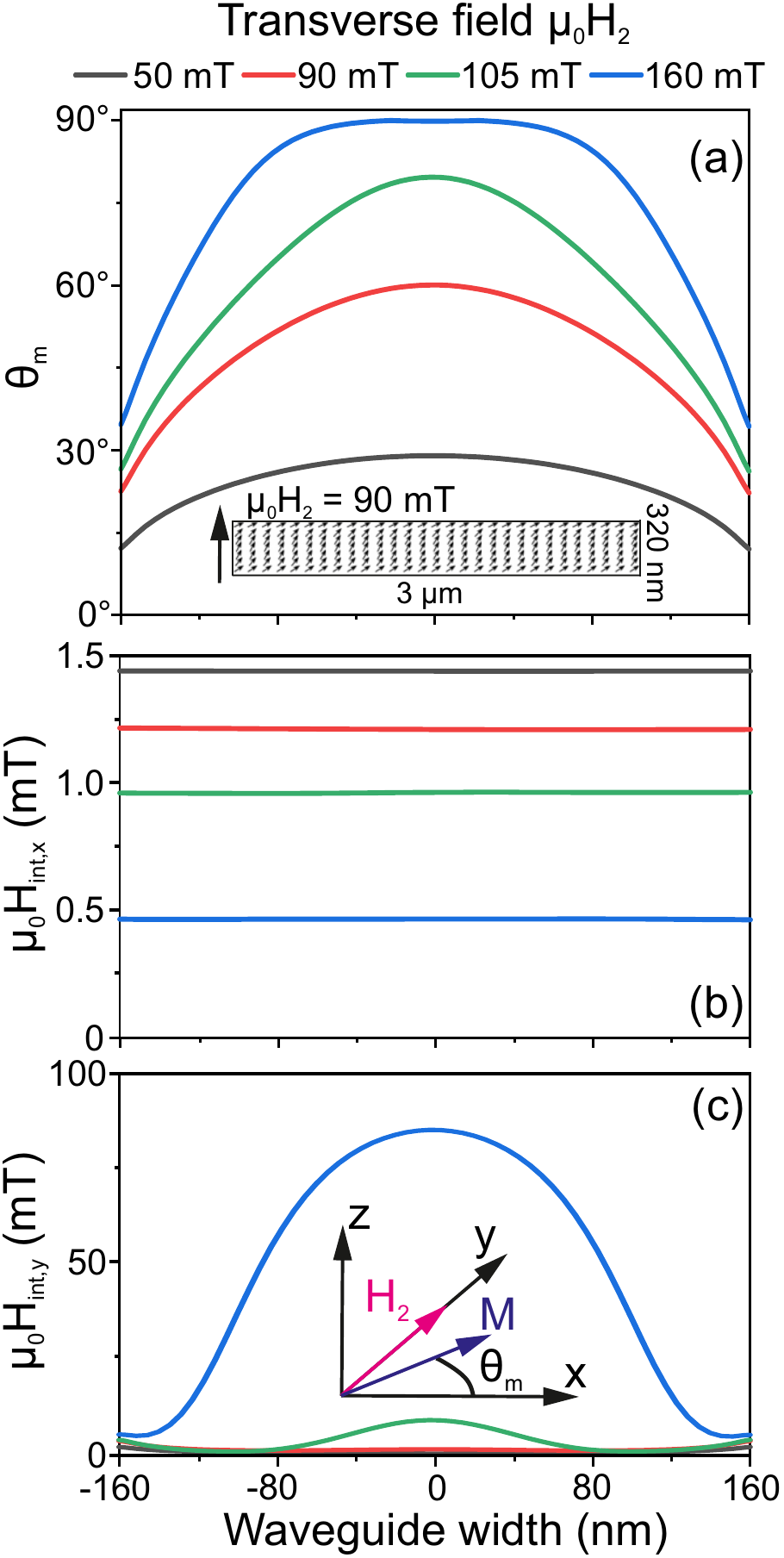}
\caption{ (a) Variation of the magnetization orientation angle $\theta_m = \arctan(m_y/m_x)$ along the width of a 320 nm wide magnetic waveguide in the center of the array for different applied bias fields. The inset image shows the static magnetization for transversal applied field $\mu_{0}H_{2}$ = 90~mT. (b) Spatial variation of the $x$- and (c) $y$-components of the internal magnetic field in the same waveguide.}
	\label{Sim_exp} 
\end{figure}

To shed further light on the magnetization dynamics and the experimental observations, micromagnetic simulations\cite{OOMMF} were performed for the narrowest waveguides (320 nm width) using the experimental geometry and magnetic material properties (exchange constant $A=18.6$~pJ/m,\cite{Talmelli_2020} $10 \times 5 \times 10$ nm$^3$ mesh). In a first step, the internal effective field and the spatial distribution of the magnetization (Fig.~\ref{Sim_exp}) in the waveguides were analyzed for different applied bias fields. The simulations found that all 20 waveguides in the array had similar distributions, with the exception of the outermost three waveguides at each side that showed slightly different magnetization and internal field orientation due to different stray fields. As a consequence of the internal field distribution, the magnetization was oriented at an average angle around or below $\theta_m = 45^{\circ}$ with respect to the $x$-axis for fields $\mu_{0}H_{2} < 105$ mT (see Fig.~\ref{Sim_exp}(a)). Only higher bias fields compensate the shape anisotropy and allow for a transveral magnetization orientation of the waveguide. Figures~\ref{Sim_exp}(b) and (c) display the $x$- and $y$-components of the internal field across the waveguide in the center of the array, respectively, for different transverse bias fields $H_2$. For bias fields $\mu_{0}H_{2} < 105$ mT, there is nearly no transverse component of the internal field since the demagnetizing field offsets the bias field. The internal fields along the waveguide ($x$-direction) are low but nonzero for small $H_{2}$ and can explain the observation of backward-volume-like spin waves.

\begin{figure*}
	\includegraphics[width=16cm]{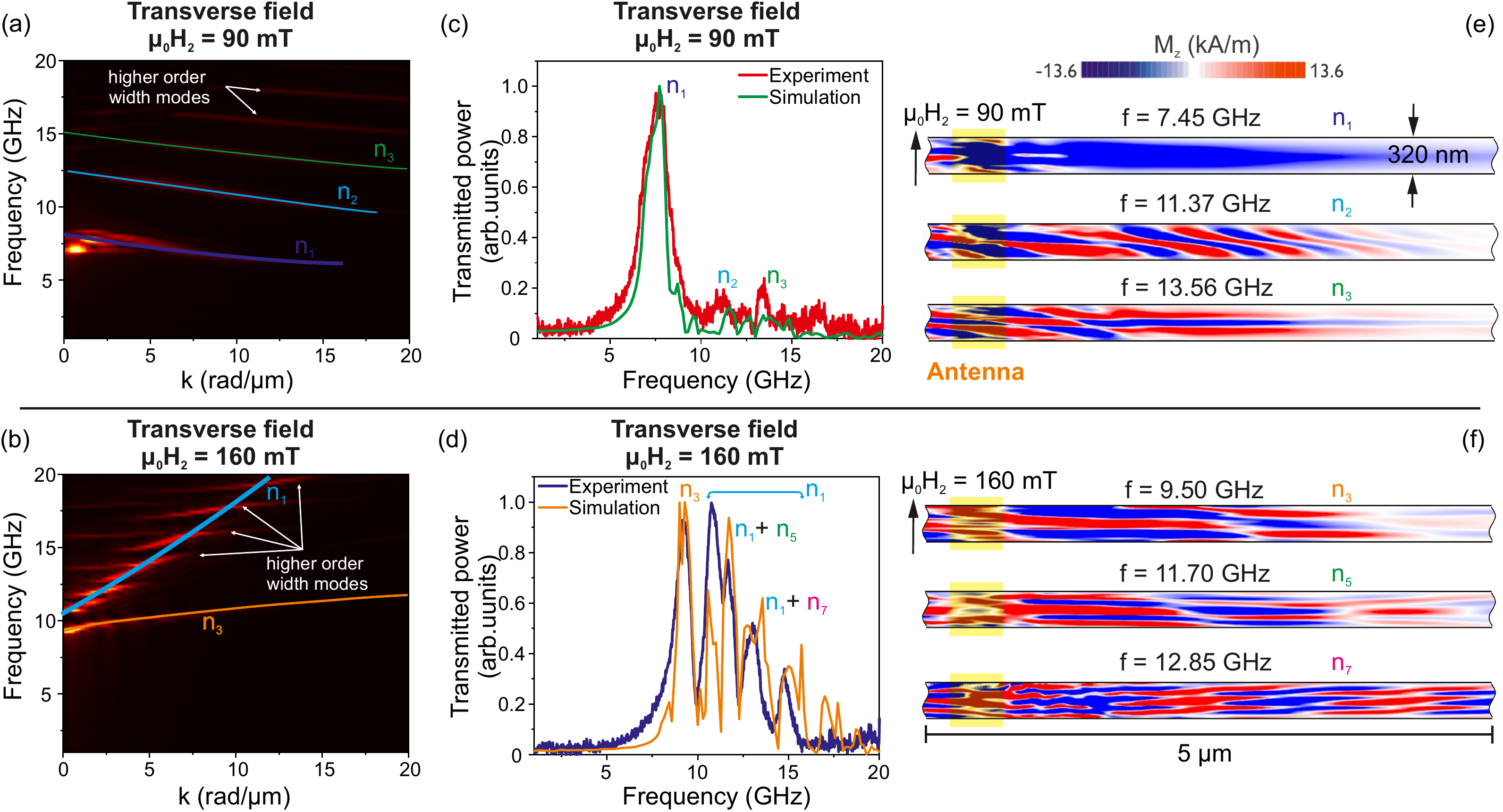}
	\caption{(a) and (b) Dispersion relations of quantized spin-wave modes extracted from micromagnetic simulations of an array of 20 magnetic waveguides (320~nm width, 180~nm gap) for applied bias fields of 90~mT and 160~mT, respectively. (c) and (d) Calculated and experimental spin-wave spectra for the two magnetic fields. (e) and (f) Snapshot images of the magnetization oscillation pattern in a magnetic waveguide in the center of the array for different excitation frequencies.}
	\label{other_exp} 
\end{figure*}

Figures \ref{other_exp}(a) and (b) depict spin-wave dispersion relations extracted from micromagnetic simulations for two representative bias fields $\mu_0 H_{2} = 90$ mT and 160~mT (\emph{i.e.} below and above $\mu_0 H_{m}$), respectively. The dispersion relations revealed several branches corresponding to different quantized width modes for both applied fields. For $\mu_0 H_{2} = 90$ mT, all branches possessed a negative slope. This behavior is reminiscent of backward volume spin waves, in agreement with an average magnetization orientation of $\theta_m<45^{\circ}$ and the analytical spin-wave dispersion relations in uniformly magnetized waveguides with an equivalent magnetization orientation. An analysis of the spatial Fourier transform of the simulated wave pattern indicated that the lowest three branches corresponded to the first- ($n_{1}$), second- ($n_{2}$) and third-order ($n_{3}$) width modes. As mentioned above, the single wire antennas used in the experiments typically cannot excite or detect even width modes in uniform waveguides due to the nearly zero overlap integral between the mode profile and the antenna Oersted field distribution.\cite{Demidov_2015} However, in the case of scaled waveguides with nonuniform tilted magnetization, the overlap integral can become nonzero\cite{Vanderveken_2020} for even modes, in agreement with the simulations in Fig.~\ref{other_exp}. 

The situation was very different for a bias field of $\mu_0 H_{2} = 160$ mT. In this case, only odd width modes appeared in the dispersion relation. All branches exhibited positive slopes, as expected for transverse magnetization. The lowest branch on the dispersion relation corresponded to the third-order width mode $n_{3}$, in qualitative agreement with analytical dispersion relations for uniform magnetization.\cite{Kalinikos_1986} By contrast, the branch corresponding to the first-order width mode $n_{1}$ intersected the dispersion relations of higher-order modes at higher frequencies. 

In a next step, the micromagnetic simulations were used to calculate the spin wave transmission spectra by averaging the magnetization dynamics over the area of the inductive antenna detector. The simulated spectra in Figs.~\ref{other_exp}(c) and (d) for bias fields of $\mu_0 H_{2} = 90$ mT and $\mu_0 H_{2} = 160$ mT, respectively, are in good agreement with the experimental results in Fig.~\ref{DE_Exp}(a) and reveal transmission peaks corresponding to different widths modes. For $\mu_0 H_{2} = 160$~mT, the transmission band corresponding to $n_{1}$ was very broad (between about 11 and 16 GHz, see Fig.~\ref{other_exp}(d)) due to the steep slope of the dispersion relation. Furthermore, its intensity was modulated due to the coherent superposition of higher-order width modes. Steady-state snapshot images of the magnetization dynamics for excitation frequencies corresponding to different width modes are shown in Figs.~\ref{other_exp}(e) and (f) for the two bias field values, respectively. The obtained magnetization profiles can be used to directly assign mode numbers to the peaks in the spin-wave transmission spectra. 

The simulations can explain well the experimental results in Fig.~\ref{DE_Exp}. For longitudinal bias fields, the behavior is similar to that confined backward volume spin waves in uniformly magnetized waveguides (\emph{cf.} Fig.~\ref{Fig_2DMaps}). However, for transverse bias fields, the behavior is strongly modified. At small $H_2$, the transverse internal field is weak and the magnetization is still oriented along the waveguide. As a result, several width modes of backward-volume-like spin waves are excited. Increasing $H_2$ leads to a decreasing effective field in the $x$-direction and decreasing spin-wave frequencies. Above a critical field $H_m$, the magnetization becomes transverse and the internal field further increases with $H_2$, leading to upward frequency shift of the spin wave band. In both regimes, higher-order width modes are also visible. Below $H_m$, a series of width modes (including even ones) is observed. Note that the critical fields at minimum frequency are band dependent and shift to higher values with increasing mode number since each mode has different quantization conditions. $H_m$ also strongly depends on the waveguide width, decreasing with increasing width (\emph{cf.} Fig.~\ref{DE_Exp}) due to reduced  shape anisotropy that allows for the saturation of wider waveguides at lower transverse bias fields. 

Above $H_m$, the lowest band corresponds to $n_3$ for the narrowest waveguide width of 320 nm; for wider waveguides, additional higher-order bands are observed below the first-order mode $n_1$, as shown in Figs.~\ref{DE_Exp}(b) to (d). This is qualitatively in agreement with the dispersion relations of spin waves in uniformly magnetized waveguides \cite{Kalinikos_1986}. Nonetheless, the strongest band corresponds in all cases to the first-order mode $n_1$. The frequency bandwidth of $n_1$ increases for wider waveguides, in keeping with the increasing slope of the dispersion relation. The $n_1$ band is modulated by the coherent superposition of higher-order modes, whose dispersion relations cross that of $n_1$, as shown in Fig.~\ref{other_exp}(b).

The results indicate that the spin-wave spectra in magnonic devices operated in the Damon-Eshbach geometry are strongly modified when the structure widths reach values of several 100 nm. Since magnonic logic devices have begun to reach such dimensions\cite{Talmelli_2020} the modified mode behavior becomes increasingly relevant for the design of next generation of devices and the analysis of their behavior. In addition, the performance of spin-wave transducers is also strongly affected since the excitation efficiency of a spin-wave mode depends on the overlap integral of its magnetization distribution with the exciting (effective) magnetic field. The results thus provide a pathway to understanding the behavior of complex spin-wave devices with dimensions at which the demagnetizing field is comparable to typical bias fields.

This work has been supported by imec’s industrial affiliate program on beyond-CMOS logic, and by the European Union's Horizon 2020 research and innovation program within the FET-OPEN project CHIRON under grant agreement No. 801055. DN and FV acknowledge financial support from the Research Foundation - Flanders (FWO) through Grants 1S89121N and 1S05719N, respectively. GT thanks Rudy Caluwaerts for SEM images, and imec’s clean room technical support.

\end{document}